\documentclass[letterpaper]{JHEP3} 
\usepackage{epsfig,array,cite,latexsym,amsmath}
\preprint{INT-PUB 03-04, UW/PT 03-04, BUHEP-03-04}
\newcommand\eqn[1]{\label{eq:#1}} 
\newcommand\Eq[1]{Eq.~\eqref{eq:#1}} 
\newcommand\half{{\textstyle{\frac{1}{2}}}} 
\newcommand\fourth{{\textstyle{\frac{1}{4}}}}

\newcommand\bfx{\mathbf{X}}
\newcommand\bfy{\mathbf{Y}}
\newcommand\bfv{\mathbf{V}}
\newcommand\bfs{\mathbf{S}}

\newcommand\bfsig{\boldsymbol{\sigma}}

\newcommand{\CC}{{\cal C}}
\newcommand{\CD}{{\cal D}}
\newcommand{\CO}{{\cal O}}
\newcommand{\CN}{{\cal N}}
\newcommand{\CP}{{\cal P}}
\newcommand{\CG}{{\cal G}}
\newcommand{\CQ}{{\cal Q}}
\newcommand{\CT}{{\cal T}}
\newcommand{\CU}{{\cal U}}

\newcommand{\CL}{{\cal L}}

\newcommand{\xh}{\mathbf{\hat{\i}}}
\newcommand{\yh}{\mathbf{\hat{\j}}}

\DeclareMathOperator{\Tr}{Tr}

\newcommand{\sla}[1]%
        {\kern .25em\raise.18ex\hbox{$/$}\kern-.55em #1}

\newcommand{\dslash}{{\boldsymbol\nabla}\cdot {\boldsymbol\gamma}\,}%
\newcommand{\mybar}[1]%
        {\kern 0.6pt\overline{\kern -0.6pt#1\kern -0.6pt}\kern 0.6pt}
\newcommand{\dig}{\kern-1.5pt \raisebox{.9ex}{$\cdot$}  \kern1.5pt
  \raisebox{0ex}{${\mathbf\cdot}$}\kern1.5pt \raisebox{-.9ex}{$\cdot$}} 
\newcommand{\digb}{\kern-1.5pt \raisebox{.75ex}{$\cdot$}  \kern1.5pt
  \raisebox{0ex}{${\mathbf\cdot}$}\kern1.5pt \raisebox{-.75ex}{$\cdot$}} 
\newcommand{\digc}{\kern-1.5pt \raisebox{1.05ex}{$\cdot$}  \kern1.5pt
  \raisebox{0ex}{${\mathbf\cdot}$}\kern1.5pt \raisebox{-1.05ex}{$\cdot$}} 
\newcommand\fverb{\setbox\pippobox=\hbox\bgroup\verb}
\newcommand\fverbdo{\egroup\medskip\noindent%
                        \fbox{\unhbox\pippobox}\ }
\newcommand\fverbit{\egroup\item[\fbox{\unhbox\pippobox}]}
\newbox\pippobox

\title{Supersymmetry on a Euclidean Spacetime Lattice I: \goodbreak A Target
  Theory with Four Supercharges}

\author{Andrew G. Cohen \\ Dept. of Physics, Boston University, 590
  Commonwealth Ave, Boston, MA  02215\\Email: \email{cohen@andy.bu.edu}}

\author{David B. Kaplan \\ Institute for Nuclear Theory, University of
  Washington, 
  Seattle, WA 98195-1550 \\Email: \email{dbkaplan@phys.washington.edu}}

\author{Emanuel Katz \\ Dept. of Physics, University
  of Washington, Seattle, WA 
  98195-1560 \\ Email: \email{amikatz@phys.washington.edu}}

\author{Mithat \"Unsal \\ Institute for Nuclear Theory, University of
  Washington, 
  Seattle, WA 98195-1550 \\Email: \email{mithat@phys.washington.edu}}

\abstract{We formulate a Euclidean spacetime lattice whose continuum
  limit is 
  $(2,2)$ supersymmetric Yang-Mills theory in two dimensions, a 
theory which possesses
  four supercharges and an anomalous global chiral symmetry.  The
  lattice action respects one exact supersymmetry, which allows the
  target theory to emerge in the continuum limit without fine-tuning.
  Our method exploits an orbifold construction described previously
  for spatial lattices in Minkowski space, and can be generalized to
  more complicated theories with additional supersymmetry and more
  spacetime dimensions.}

\begin{document} 

\section{Introduction and results}
\label{sec:1}

Supersymmetric Yang-Mills (SYM) theories are extraordinarily rich,
exhibiting a variety of complex nonperturbative phenomena.  Much is
known analytically about such theories, yet presumably much more could
be learned from numerical study.  Lattice versions of supersymmetric
field theories have been discussed at length in the literature
\cite{Dondi:1977tx,Banks:1982ut,Elitzur:1982vh,Bartels:1983wm,%
Golterman:1989ta,Montvay:1996ea,Nishimura:1997vg,Neuberger:1998bg,%
Nishimura:1998hu,Bietenholz:1998qq,Kaplan:1999jn,Fleming:2000fa,%
Fleming:2000hf,Farchioni:2001wx,Catterall:2001wx,Catterall:2001fr,%
Catterall:2003wd,Itoh:2002nq,Fujikawa:2002ic,Fendley:2002sg} (for a
recent review see \cite{Feo:2002yi}).  A persistent obstacle to
numerical simulation of supersymmetric theories is the lack of
continuous translational invariance, which makes exact realization of
the desired supersymmetry on the lattice impossible.  Without this
exact supersymmetry, fine-tuning seems the only approach to obtaining
the supersymmetric continuum limit.  The problem is especially acute
for the most interesting case of non-minimal SYM theories, which
contain spin zero interacting fields: without exact lattice
supersymmetry it would seem that fine-tuning of the bare lattice
parameters would be needed to eliminate radiative corrections to
scalar masses and interactions.

A method to circumvent this problem was recently proposed for spatial
lattices with continuous Minkowski time \cite{Kaplan:2002wv}. The
lattices are obtained by an orbifold projection of a ``mother theory''
that has as much supersymmetry as the target theory.  The projection
leaves a subset of the supersymmetries realized exactly on the
lattice, protecting the theory from unwanted relevant operators in the
continuum limit\footnote{For other recent work on maintaining exact
  supersymmetries on the lattice, see
  \cite{Catterall:2001fr,Catterall:2001wx,Catterall:2003wd,Itoh:2002nq}.}.
It was shown that the desired continuum theory, with its enhanced
super and chiral symmetries, can be obtained without fine tuning.
This procedure, inspired by recent work on orbifolding
\cite{Douglas:1996sw,Kachru:1998ys,Schmaltz:1998bg,Strassler:2001fs}
and the deconstruction of supersymmetric theories
\cite{Arkani-Hamed:2001ca,Arkani-Hamed:2001ie,Csaki:2001em,Rothstein:2001tu},
leads to rather unorthodox lattices, with spin zero bosons and single
component fermions living on both links and sites, as well as
noncompact gauge fields. The lattice structures also tend to be more
exotic than the $d$-cubic lattices normally considered in $d$
dimensions\footnote{For related work on nonsupersymmetric theories,
  see \cite{Hill:2000mu,Cheng:2001vd,Hill:2002me}.}.

This is the first of several papers in which we extend the previous
work to Euclidean spacetime lattices, which are more suitable for
numerical simulation.  Here we work through a simple example in some
detail: the lattice for $(2,2)$ SYM theory in two spacetime dimensions
with $U(k)$ gauge symmetry\footnote{This model has been studied
  numerically using Discrete Light-Cone Quantization techniques in
  Ref.~\cite{Antonuccio:1998mq}}.  This target theory is described by
the Lagrangian
\begin{equation} 
  \CL =  \frac{1}{g_2^2}  \, \Tr
  \Biggl(\bigl\vert D_m s\bigr\vert^2 + \mybar \psi \,i  D_m
  \gamma_m 
  \psi + \fourth v_{mn} v_{mn} +i\sqrt{2}(\mybar \psi_L [ s, \psi_R]
  +\mybar\psi_R [ s^\dagger, \psi_L]) + \half
  [s^\dagger,s\,]^2\Biggr) 
  \eqn{targ2}
\end{equation} 
where $\psi_R$ and $\psi_L$ are the right- and left-chiral components
of a two-component Dirac field $\psi$, $s$ is a complex scalar field,
$v_m$ is the two dimensional gauge potential, $D_m = \partial_m + i
[v_m,\;\cdot\;]$ is the covariant derivative, and $v_{mn} =-
i[D_m,D_n]$ is the field strength. All fields are rank-$k$ matrices
transforming as the adjoint representation of $U(k)$. This target
theory has four supercharges and a chiral $U(1)$ $R$-symmetry, and can
be obtained by dimensionally reducing the more familiar ${\cal N}=1$
SYM theory in four dimensions down to two dimensions.

\begin{figure}[t]
\centerline{\epsfxsize=7cm\epsfbox{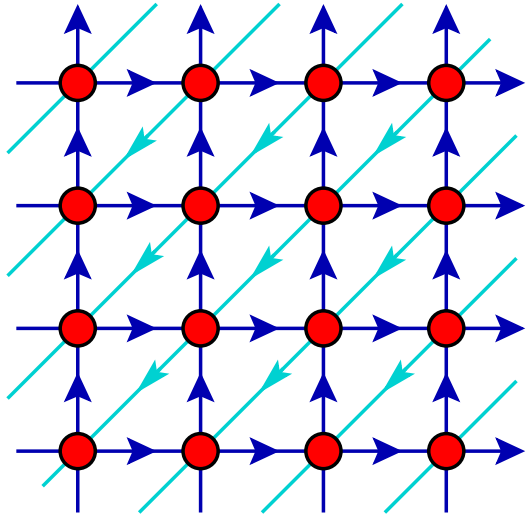}}
\smallskip
\caption{\sl The two dimensional Euclidean lattice for the two
  dimensional SYM theory with four supercharges.  The lattice respects one exact
  supercharge. Arrows on the links signify the orientation of
   the $\mathbf{r}$ charge vectors of the link fermions, as discussed
   in \S2.}
  \label{fig:fig1} 
\end{figure}

A na\"\i{}ve latticization of this model breaks both the supersymmetry
and the chiral $U(1)$ symmetry, requiring fine tuning to eliminate
radiatively induced operators which violate these symmetries.  Here we
construct a supersymmetric lattice action which does not suffer from
these problems, exhibiting both supersymmetry and chiral symmetry in
the continuum without the need for fine-tuning of parameters.  Our
resulting lattice action is
\begin{multline}
  S=\frac{1}{g^2} \sum_{\mathbf n} \Tr\Bigl[\half \left(\mybar
    x_{\mathbf{n}-\xh} x_{\mathbf{n}-\xh} - x_{\mathbf n}\mybar
    x_{\mathbf n}+ \mybar y_{\mathbf{n}-\yh}
    y_{\mathbf{n}-\yh} - y_{\mathbf n}\mybar y_{\mathbf
      n}\right)^2 +2\,\bigl| x_{\mathbf n} y_{\mathbf{n} +\xh} -
  y_{\mathbf n} x_{\mathbf{n} + \yh}\bigr|^2
  \Bigr.\\
  +\sqrt{2}\,\left(\alpha_{{\mathbf n}} \mybar x_{\mathbf n
    }\lambda_{{\mathbf n}} - \alpha_{\mathbf{n}-\xh}
    \lambda_{\mathbf{n}} \mybar x_{\mathbf{n}-\xh}\right)
    +\sqrt{2}\,\left(\beta_{{\mathbf n}} \mybar 
    y_{\mathbf n }\lambda_{\mathbf{n}} - \beta_{\mathbf{n}-\yh}
    \lambda_{\mathbf{n}} \mybar y_{\mathbf{n}-\yh}\right)\\ 
  \Bigl.  -\sqrt{2}\,\left(\alpha_{\mathbf{n}} 
    y_{\mathbf{n}+\xh} \xi_{\mathbf{n}} -
    \alpha_{\mathbf{n}+\yh} \xi_{\mathbf{n}} y_{\mathbf n}\right)
  +\sqrt{2}\,\left(\beta_{\mathbf{n}} x_{\mathbf{n}+\yh}\xi_{\mathbf{n}} -
    \beta_{\mathbf{n}+\xh} \xi_{\mathbf{n}} x_{\mathbf n}\right)\\
  +a^2\mu^2\left( x_{\mathbf{n}}\mybar x_{\mathbf{n}}
    -\frac{1}{2a^2}\right)^2 + a^2\mu^2 \left( y_{\mathbf{n}}\mybar
    y_{\mathbf{n}} -\frac{1}{2a^2}\right)^2\Bigr]
  \eqn{latact}
\end{multline}
Here the vectors $\mathbf{n}\equiv\{n_x,n_y\}$ label the sites of the
2-dimensional $N\times N$ periodic lattice pictured in
Figure~\ref{fig:fig1}.  All variables are $k\times k$ matrices; Latin
variables are bosonic, while Greek variables are Grassmann.  The
variables $x, \mybar x \text{ and } \alpha$ reside on the links in the
$\xh=\{1,0\}$ direction, $y, \mybar y \text{ and } \beta$ are
variables on the links in the $\yh=\{0,1\}$ direction, $\xi$ sits on
the diagonal links, and $\lambda$ is a site variable. The parameter
$a$ appearing in the action has dimension of length, and is
interpreted as the lattice spacing.  The continuum limit is defined as
$a \to 0, N \to \infty$ while holding the 2-dimensional coupling $g_2
\equiv g a$ and the lattice size $L \equiv N a$ fixed.
%
%

The term proportional to $\mu^2$ softly breaks the exact
supersymmetry, and in the continuum limit $\mu$ is scaled such that
$g_2 a \ll \mu L\lesssim 1$.  The soft supersymmetry breaking effects
then vanish in the large volume limit.

The remainder of this paper examines the route which leads to the
peculiar lattice action \Eq{latact}.  The lattice arises from an
orbifold projection of a matrix model with four exact supercharges.
We examine the symmetries of the resulting lattice action and show
that for $\mu=0$ it preserves one exact supersymmetry.  We demonstrate
that the continuum limit of the lattice is the SYM theory of
\Eq{targ2}, and that the exact lattice supersymmetry obviates the need
for fine tuning in the quantum theory. As an aside, note that the
lattice \Eq{latact} is an example of an interacting theory for which
the critical point for obtaining chiral symmetry in the continuum
limit is exactly known.

\section{Constructing the lattice action}
\label{sec:2}

Our technique for obtaining a Euclidean spacetime lattice begins with
an $\CN=1$ SYM theory with $U(k N^2)$ gauge group and $\CQ=4$
supercharges in $d=4$ (Euclidean) dimensions\footnote{Euclidean
  supersymmetry has been discussed in detail in
  ref.~\cite{Nicolai:1978vc}.}. This theory is then dimensionally
reduced to zero dimensions, yielding a matrix model containing both
bosonic and fermionic rank-$k N^2$ matrices transforming as $\Phi \to
g \Phi g^{-1}$ under the symmetry $\CG=U(kN^2)$.  The action of this
``mother theory'' is given by
\begin{equation}
  S = \frac{1}{g^2}\left(\frac{1}{4} \Tr v_{mn} v_{mn} + \Tr
    \mybar\psi\,  \mybar\sigma_m [v_m, 
    \psi]\right)\ ,
  \eqn{momiv}
\end{equation}
where $m,n=0,\ldots,3$, $\psi$ and $\mybar\psi$ are independent
complex two-component spinors, $v_m$ is the 4-vector of gauge
potentials, and
\begin{equation}
  v_{mn} = i[v_m,v_n]\ , \quad \sigma_m = \{ 1,\, -i{\bfsig}\}\  ,\quad
  \mybar\sigma_m = \{ 1,\, i\bfsig\}\ ,
\end{equation}
The ``daughter theory'' we will derive from this by orbifolding will
be a two-dimensional lattice with $N^2$ sites and a $U(k)$ symmetry
associated with each site.

This mother action possesses an $R$-symmetry $G_R=SU(2)\times
SU(2)\times U(1)$.  The $SU(2)\times SU(2)\simeq SO(4)$ subgroup of
$G_R$ is the Lorentz symmetry of the Euclidean theory prior to
dimensional reduction, under which
\begin{equation}
  v \equiv v_m\mybar\sigma_m \to L v R^\dagger\ ,\qquad \mybar\psi\to
  \mybar\psi L^\dagger\ ,\qquad \psi \to R \psi\ ,
  \eqn{lrtrans}
\end{equation}
where $L$ and $R$ are independent $SU(2)$ matrices.  The $U(1)$ factor
of $G_R$ is just the chiral symmetry of the gauginos under which
\begin{equation}
  \psi \to e^{i\theta/2} \psi\ ,\qquad \mybar\psi\to \mybar\psi
  e^{-i\theta/2}\ ,\qquad v_m\to v_m\ .
  \eqn{qtrans}
\end{equation}
We will denote the seven generators of the group $G_R=SU(2)\times
SU(2)\times U(1)$ by
\begin{equation}
  L_a\,,\ R_a\,,\ Y\ ,
\end{equation}
where $L_a$ and $R_a$ are $SU(2)$ generators and the $U(1)$ generator
$Y$ is normalized so that $\psi$ and $\mybar\psi$ carry charges
$Y=\pm\half$.

In addition to this $R$-symmetry the mother theory \Eq{momiv} is also
invariant under four independent supersymmetry transformations
parametrized by two-component Grassmann spinors $\kappa$ and
$\mybar\kappa$:
\begin{gather}
  \delta v_m = -i\mybar \psi\, \mybar\sigma_m\kappa + i \mybar\kappa \,\mybar
  \sigma_m\psi\ ,\quad
  \delta \psi = -i v_{mn}\sigma_{mn}\kappa \ ,\quad
  \delta \mybar \psi = i v_{mn} \,\mybar\kappa \,\mybar\sigma_{mn}\ ,
  \eqn{trans}
\intertext{where}
  \sigma_{mn} \equiv
  {\textstyle{\frac{i}{4}}}
  \left(\sigma_m\mybar\sigma_n-\sigma_n\mybar\sigma_m\right)\ ,\qquad\qquad
  \mybar\sigma_{mn} \equiv
  {\textstyle{\frac{i}{4}}}
  \left(\mybar\sigma_m\sigma_n-\mybar\sigma_n\sigma_m\right)\ . 
\end{gather}

A two dimensional lattice with $N^2$ sites may be created out of the
rank-$k N^2$ matrices of the mother theory, following the procedure in
\cite{Kaplan:2002wv}. We define a $Z_N\times Z_N \equiv \Gamma$
subgroup of $\CG\times G_R$ and project out all field components not
invariant under $\Gamma$.  The embedding of $\Gamma$ is designed to
produce a lattice with nearest and next-to-nearest neighbor
interactions, and one exact supersymmetry.  The generators of the two
$Z_N$ factors in $\Gamma$ are denoted as $\hat{\gamma}_1$,
$\hat{\gamma}_2$, and their action on a field $\Phi$ of the mother theory is
\begin{equation}
  \hat{\gamma}_a \Phi = e^{2\pi i r_a/N} \CC^{(a)} \Phi {\CC^{(a)}}^{-1}\
  \qquad \text{with}\quad a=1,2\ 
  \eqn{orbi}
\end{equation}
where $e^{2\pi i r_a/N}\in G_R$ and $\CC^{(a)}\in \CG$.  The
$\CC^{(a)}$ are referred to as ``clock'' matrices.  It is convenient
to write our rank-$k N^2$ matrix field $\Phi$ in the form
$\Phi_{\mu\nu i,\, \mu' \nu' j}$ with Greek letters running from $1$
to $N$, and Latin letters running from $1$ to $k$. The clock matrix
$\CC^{(1)}$ in \Eq{orbi} is taken to act nontrivially on the $\mu$,
$\mu'$ indices, while $\CC^{(2)}$ acts on $\nu$, $\nu'$.  In this
notation, the $\CC^{(a)}$ are the $\CG=U(kN^2)$ matrices
\begin{equation}
  \CC^{(1)} = \Omega\otimes {\mathbf 1}_N \otimes  {\mathbf 1}_k\ ,\qquad
  \CC^{(2)} =  {\mathbf 1}_N\otimes \Omega \otimes  {\mathbf 1}_k\ ,
\eqn{clocka}
\end{equation}
where ${\mathbf 1}_d$ signifies a rank-$d$ unit matrix, and $\Omega$ is
the rank-$N$ unitary matrix
\begin{equation}
  \Omega = 
  \begin{pmatrix} 
    \  \, \omega\  \,  & & &\\ &\  \,
    \omega^2\,  \ & & \cr &&\dig &\\
    &&&\omega^{N} 
  \end{pmatrix}
  \qquad \text{with} \quad \omega\equiv e^{2\pi i/N}\ .
\eqn{clockb}\end{equation}
The integer charges $\mathbf{r}=\{r_1,r_2\}$ in \Eq{orbi} are
constructed from the Cartan sub-algebra of $G_R= SU_L(2)\times
SU_R(2)\times U_Y(1)$:
\begin{equation}
  r_1 = -L_3+R_3- Y\ ,\quad r_2 = +L_3+R_3- Y\ .
  \eqn{rdef}
\end{equation}

The orbifold projection eliminates the components of the mother theory
fields which are not invariant under the $\Gamma$ transformation
defined in \Eq{orbi}.  The projection operator may be written as
\begin{equation}
  \hat{P} = \frac{1}{N^2}\sum_{m,n=1}^N\
  (\hat{\gamma}_1)^m(\hat{\gamma}_2)^n\ .
\end{equation}
The ``daughter'' theory is obtained by replacing every field $\Phi$ in
the action of the mother theory by its projection $\tilde{\Phi} =
\hat{P} \Phi$.  Each projected field $\tilde{\Phi}$ consists of $N^2$
nonzero $k\times k$ blocks; the position of these nonzero blocks
within the original rank-$k\,N^2$ matrix is determined by the integer
values of the $r_{1,2}$ charges of $\Phi$.  Choosing a basis for the
mother theory fields such that the $r_a$ charges are diagonal, the
nonzero blocks in $\tilde\Phi_{\mu\nu i,\mu'\nu' j}$ occur only for
\begin{equation}
  \mu' = \mu + r_1\ ,\qquad \nu'=\nu+r_2\ ,
  \eqn{rmean}
\end{equation}
where $r_1$ and  $r_2$ refer to the particular charges carried by that field $\Phi$.
Each nonzero rank-$k$ block in $\tilde{\Phi}$ is labeled by the
quartet of indices $\{\mu,\nu\},\{\mu',\nu'\}$. We can consider these blocks as
lattice variables of a two dimensional lattice, with each lattice site
labeled by the pair $\{\mu,\nu\}$.  Each block is then a link
variable, residing on the link between sites $\{\mu,\nu\}$ and
$\{\mu',\nu'\}$---or residing at site $\{\mu,\nu\}$ if $\mu=\mu'$ and
$\nu=\nu'$.  We see from \Eq{rmean} that fields with
$\mathbf{r}=\{0,0\}$ are site variables, while those with non-zero
$\mathbf{r}$ are variables for links pointing in the $\mathbf{r}$ direction.

The variables of the mother theory are easily expressed in
terms of eigenstates of the $\mathbf{r}$ charges:
\begin{gather}
  x = \frac{v_0 - i v_3}{\sqrt{2}}\ ,\qquad 
  \mybar x = \frac{v_0 + i v_3}{\sqrt{2}}\ ,\qquad 
  y= -i\frac{v_1 - i
    v_2}{\sqrt{2}}\ ,\qquad\mybar y= i\frac{v_1 + i
    v_2}{\sqrt{2}}\ .
\intertext{and}
  \psi =
  \begin{pmatrix}
    \lambda\\ \xi\\ 
  \end{pmatrix} \ , \qquad
  \mybar\psi =
  \begin{pmatrix}
    \alpha & \beta \\
  \end{pmatrix}\ .
\end{gather}

The charge assignments for each of the variables in this action are
shown in Table~\ref{tab:tab1}.  They follow from the transformations
given in \Eq{lrtrans} and \Eq{qtrans}, along with the definition of
the $r_a$ in \Eq{rdef}. The daughter theory lattice we obtain after
orbifold projection is shown in Figure~\ref{fig:fig1}.
In terms of these rank-$k N^2$ matrices, the action of the mother
theory \Eq{momiv} may be written as
\begin{equation} 
  S=\frac{1}{g^2}\Tr\left( \half\left([\mybar x, x] +
      [\mybar y, y]\right)^2 +2\,\Bigl|[ x, y]\Bigr|^2
    +\sqrt{2}\, \left(\alpha[\mybar x,\lambda] +\beta
      [\mybar y,\lambda] -\alpha 
      [ y,\xi] +\beta [ x, \xi]\right)
  \right)
  \eqn{act4}
\end{equation}

\setlength{\extrarowheight}{5pt}
\begin{table}[t]
\centerline{
\begin{tabular}
{|c||c|c|c|c||c|c|c|c|}
\hline
&$  x $&$\mybar  x $&$ y$&$\mybar  y$&$ \lambda$&$
\xi$&$\alpha$&$\beta$
\\ \hline
$L_3 $&$ -\half $&$ +\half $&$ +\half $&$ -\half $&$  \,\ 0 $&$ \,\ 0 $&$
 -\half $&$ +\half $
\\ 
$R_3 $&$ +\half $&$ -\half $&$ +\half $&$ -\half $&$  +\half $&$
-\half $&$   \,\ 0 $&$  \,\ 0 $
\\    
$ Y $ &$  \,\ 0 $&$ \,\ 0 $&$ \,\ 0 $&$ \,\ 0 $&$ +\half $&$ +\half
$&$ -\half $&$ -\half $
\\ \hline
$ r_1 $&$ +1 $&$ -1 $&$ \,\ 0 $&$ \,\ 0 $&$ \,\ 0 $&$ -1 $&$  +1
$&$\,\ 0 $
\\
$ r_2 $&$\,\ 0 $&$ \,\ 0 $&$ +1 $&$ -1 $&$ \,\ 0 $&$ -1 $&$ \,\ 0 $&$
+1 $
\\ \hline
 \end{tabular}
}
\caption{\sl The  $r_{1,2}$ charges  of the  fields of the $\CQ=4$ mother 
 theory
which
  define the orbifold projection.\label{tab:tab1}}
\end{table}

The location of each variable on the lattice can be read off from the
corresponding $\mathbf{r}$ charges given in Table~\ref{tab:tab1}: the
$\lambda$ fermion lives at the sites, while $\xi$ lives along the
diagonal links; $x$, $\mybar x$ and $\alpha$ reside along the
$\xh$-links; $y$, $\mybar y$ and $\beta$ reside along the $\yh$-links.
The lattice action is found by replacing all of the matrix variables
in the action of the mother theory \Eq{act4} by their orbifold
projection\footnote{We label the coordinates of our variables such
  that $\lambda_{\mathbf n}$ resides at site ${\mathbf n}$;
  $x_{\mathbf n}$, $\mybar x_{\mathbf n}$ and $\alpha_{\mathbf n}$
  live on the $(\mathbf{n},\mathbf{n}+\xh)$ link; $y_{\mathbf n}$,
  $\mybar y_{\mathbf n}$ and $\beta_{\mathbf n}$ live on the
  $(\mathbf{n},\mathbf{n}+\yh)$ link; and $\xi_{\mathbf n}$ sits on
  the $(\mathbf{n},\mathbf{n} + \xh + \yh)$ diagonal links.  Link
  variables have an orientation determined by their vector
  $\mathbf{r}$.  Shown in Figure~\ref{fig:fig1} are the orientations
  of the link fermions, $\alpha$, $\beta$ and $\xi$.  The bosonic link
  variables $x$ and $y$ are oriented in the same sense as $\alpha$ and
  $\beta$ respectively, while $\mybar x$ and $\mybar y$ are
  anti-oriented.}
\begin{multline}
  S=\frac{1}{g^2} \sum_{\mathbf n} \Tr\Bigl[\half \left(\mybar 
    x_{\mathbf{n}-\xh} x_{\mathbf{n}-\xh} - x_{\mathbf
    n}\mybar x_{\mathbf n}+ 
    \mybar y_{\mathbf{n}-\yh} y_{\mathbf{n}-\yh} - y_{\mathbf n}\mybar
    y_{\mathbf n}\right)^2 +2\,\bigl| x_{\mathbf n} y_{\mathbf{n} + \xh}
  - y_{\mathbf n} x_{\mathbf{n} + \yh}\bigr|^2
  \Bigr.\\
  +\sqrt{2}\,\left(\alpha_{{\mathbf n}} \mybar x_{\mathbf n
    }\lambda_{{\mathbf n}} - \alpha_{\mathbf{n}- \xh}\lambda_{{\mathbf
        n}} \mybar x_{\mathbf{n}- \xh}\right)
  +\sqrt{2}\,\left(\beta_{{\mathbf n}} \mybar y_{\mathbf n
    }\lambda_{{\mathbf n}} - \beta_{\mathbf{n}- \yh}\lambda_{{\mathbf
        n}} \mybar y_{\mathbf{n}- \yh}\right)\\ \Bigl.
  -\sqrt{2}\,\left(\alpha_{{\mathbf n }} y_{\mathbf{n}+\xh}\xi_{{\mathbf
        n}} - \alpha_{\mathbf{n}+\yh}\xi_{{\mathbf n}} y_{\mathbf
      n}\right) +\sqrt{2}\,\left(\beta_{{\mathbf n}} x_{\mathbf{n}
      +\yh}\xi_{{\mathbf n}} - \beta_{\mathbf{n}+\xh}\xi_{{\mathbf
        n}} x_{\mathbf n}\right)\Bigr] \eqn{q4lat}
\end{multline}
where the vectors $\mathbf{n}=\{n_x,n_y\}$ label the lattice sites on
the $N\times N$ periodic lattice of Fig.~\ref{fig:fig1}.

\section{Symmetries of the lattice action}
\label{sec:3}

The seemingly elaborate construction of the above lattice action is
warranted by the symmetry properties it possesses---symmetries which
ensure that the target theory arises in the continuum limit without
fine tuning of operators.  The symmetries of the lattice action
include all symmetries of the mother theory which commute with the
orbifold projection operator $\hat{P}$, defined previously as
\begin{equation}
  \hat{P} = \frac{1}{N^2}\sum_{m,n=1}^N\
  (\hat{\gamma}_1)^m(\hat{\gamma}_2)^n\ ,\qquad 
  \hat{\gamma}_a \Phi = e^{2\pi i r_a/N} \CC^{(a)} \Phi
  {\CC^{(a)}}^{-1}  \qquad \text{with} \quad a=1,2\ .
  \eqn{proj}
\end{equation}
These consist of a $U(k)$ gauge symmetry ({\it e.g.\/} a $ U(k)^{N^2}$
 symmetry, with an independent $U(k)$ symmetry associated with
each site); a $Z_N\times Z_N$ discrete translational symmetry; a $U(1)^3$
global internal symmetry; a $Z_2$ lattice point group consisting of
reflections about the diagonal link; and a $\CQ=1$ supersymmetry.
These symmetries are important for determining the renormalization
properties of the theory, and so we demonstrate how each of these
symmetries commutes with the orbifold projection and describe the
action on component fields.  We pay particular attention to the
supersymmetry, and reexpress the lattice action \Eq{q4lat} in terms of
superfields so that the possible form of counterterms becomes
transparent.

\subsection{The $U(k)$ gauge symmetry}
\label{sec:4a}
The $U(k)$ gauge symmetry corresponds to an independent $U(k)$
symmetry associated with each of the $N^2$ lattice sites.  This
$U(k)^{N^2}$ symmetry is a subgroup of the $\CG=U(kN^2)$ symmetry of
the mother theory.  Using the direct product notation $\Phi_{\mu\nu
  i,\mu'\nu' j}$, where $\mu$, $\nu$, $\mu'$, $\nu'$ run from
$1,\ldots,N$ and $i$, $j$ run from $1,\ldots, k$, for any of the
$kN^2\times kN^2$ matrices of the mother theory, the action of the
$U(k)^{N^2}$ symmetry may be written as
\begin{equation}
  \Phi \to \CU \Phi \CU^\dagger\ ,\qquad \CU_{\mu\nu i, \mu'\nu' j} = 
  U^{(\mu\nu)}_{ij}\delta_{\mu \mu'} \delta_{\nu \nu'}\ ,
\end{equation}
where $U^{(\mu\nu)}$ is an independent rank-$k$ unitary matrix for
each of the $N^2$ values of $(\mu,\nu)$.  This transformation commutes
with the orbifold projection \Eq{proj}:  it acts nontrivially
only in the $k\times k$ subspace corresponding to indices $i$, $j$,
while the clock matrices $\CC^{(a)}$ act nontrivially only on the
$\mu$, $\nu$ indices
\begin{equation}
  \CC^{(1)}_{\mu\nu i, \mu'\nu' j} = \omega^{\mu} \delta_{\mu\mu'}
  \delta_{\nu \nu'}\delta_{i j}\ ,\qquad
  \CC^{(2)}_{\mu\nu i, \mu'\nu' j} = \omega^{\nu} \delta_{\mu\mu'}
  \delta_{\nu \nu'}\delta_{i j}\qquad \text{with}\quad \omega\equiv
  e^{2\pi i/N}\ . 
\end{equation}
This $U(k)$ gauge symmetry also commutes with the $r_a$
charges that appear in the projection operator $\hat{P}$, as they live
in a different space.

A site variable on the lattice such as $\lambda_{\mathbf n}$
transforms as an adjoint under the $U^{({\mathbf n})}(k)$ symmetry
associated with that site; link variables such as $x_{\mathbf n}$ or
$\mybar y_{\mathbf n}$ transform as bifundamentals under the two
$U(k)$ symmetries associated with the sites at the link's endpoints.
The orientation of a link variable is determined by its $\mathbf{r}$
charges.  For example, since $x_{\mathbf n}$ has $\mathbf{r}=\{1,0\}$,
it transforms as $x_{\mathbf n} \to U^{({\mathbf n})}x_{\mathbf n}
U^{(\mathbf{n}+\xh)\,\dagger}$, while $\mybar y_{\mathbf n}$ has
$\mathbf{r}=(0,-1)$ and so transforms as $\mybar y_{\mathbf n} \to
U^{(\mathbf{n}+\yh)}\mybar y_{\mathbf n} U^{({\mathbf n})\,\dagger}$.
Gauge invariant operators are constructed from traces of products of
variables, corresponding to oriented closed loops on the lattice.

\subsection{The $Z_N\times Z_N$ translation symmetry}
\label{sec:4b}
The translation symmetries of the lattice also arise from $\CG$
symmetries of the mother theory.  We define the ``shift'' matrices
\begin{gather}
  \CT^{(1)} = T\otimes  {\mathbf 1}_N \otimes  {\mathbf 1}_k\ ,\qquad
  \CT^{(2)} =  {\mathbf 1}_N\otimes 
  T \otimes  {\mathbf 1}_k\ , 
\intertext{where}
  T = 
  \begin{pmatrix}
    0\ \  &\ \ 1\ \ &   &       \\
    & \,\,\digc &\,\, \digc &   \\
    &  &    \ \ 0\ \ &\ \ 1\  \\
    1\ \ & &  &\ \ 0\ \\
  \end{pmatrix}
\eqn{shifta}
\end{gather}
The two shift operators $\hat{t}_a$ are defined to act on the fields of
the mother theory as $\hat{t}_a \Phi = \CT^{(a)} \Phi{\CT^{(a)}}^{-1}$,
and they generate a $Z_N\times Z_N$ subgroup of $\CG$.  These shift
matrices commute with the $r_a$ charges, while conjugation of the
clock matrices by the shift matrices is simply
\begin{equation}
  \CT^{(a)} \CC^{(b)} {\CT^{(a)}}^{-1} = 
  \begin{cases}
    \omega\,
    \CC^{(b)}\ , & 
    a=b\\ \CC^{(b)}\ , &
    a\ne b\ ,
  \end{cases}
\eqn{shiftb}
\end{equation}
It follows that the shift operators $\hat{t}_a$ commute with the
$Z_N\times Z_N$ generators $\hat{\gamma}_b$, and hence with the orbifold
projection operator $\hat{P}$:
\begin{equation}
  [\hat{t}_a, \hat{\gamma}_b]= [\hat{t}_a,\hat{P}]=0 \qquad \text{with}
  \quad a,b=1,2\ .
\end{equation}
Since the $\hat{t}_a$ commute with $\hat{P}$, they generate symmetries
of the lattice action.  By considering the action of the shift
matrices $\CT^{(a)}$ on the lattice component fields, one can see that
the $\hat{t}_a$ generate the $Z_N\times Z_N$ lattice translations, with
$\hat{t}_1$ generating translations in the $\xh$ direction,
and $\hat{t}_2$ generating translations in the $\yh$
direction.  Note that the $\hat{t}_a$ as defined do not commute with
the $U(k)$ gauge symmetry.

\subsection{The $U(1)^3$ global symmetry}
\label{sec:4c}

The $U(1)^3$ global symmetry is generated by $r_1$, $r_2$, and $Y$.
These charges live in the Cartan sub-algebra of the $G_R$ symmetry
group of the mother theory, and so commute with the $\CC^{(a)}\in\CG$
in \Eq{proj}.  They also clearly commute with the $r_a\in G_R$, which
we chose to equal $\pm L_3 + R_3-Y$. Since gauge invariant operators
correspond to closed loops on the lattice, and since $r_1$ and $r_2$
specify the link vectors, each gauge invariant operator will
necessarily have net charges $r_1=r_2=0$. The remaining $U(1)$
generated by $Y$ is nontrivial, however, and corresponds to an exact
$U(1)$ $R$-symmetry on the lattice, which in turn becomes the exact
$U(1)$ fermion number symmetry of the target theory in \Eq{targ2}.

\subsection{The $Z_2$ point group}
\label{sec:4d}

The lattice in Fig.~\ref{fig:fig1} respects a $Z_2$ symmetry,
corresponding to reflections about the diagonal link. The generator
$\hat{\sigma}_d$ of this
symmetry resides in both the $G_R$ and the $\CG$ groups of the mother
theory.  Its action on a field $\Phi$ of the mother theory is
\begin{equation}
  \hat{\sigma}_d\, \Phi = \hat{g} \,\CD \Phi \CD^{-1}\
  \quad\text{with}\quad \hat{g} = e^{i\pi( L_1+R_3 +Y)}\in G_R\ ,\quad
  \CD_{\mu\nu 
    i,\mu'\nu'j}=\delta_{\mu\nu'} \delta_{\nu\mu'}\delta_{ij}\in \CG
\end{equation}
These matrices have the properties
\begin{subequations}
\begin{align}
    \hat{g}\, e^{2\pi i r_1/N}\hat{g}^{-1} &=  e^{2\pi i r_2/N}
    & \CD\, \CC^{(1)} \CD^{-1} &= \CC^{(2)}\\
    \hat{g}\, e^{2\pi i r_2/N}\hat{g}^{-1} &=  e^{2\pi i r_1/N}
    & \CD\, \CC^{(2)} \CD^{-1} &= \CC^{(1)}\ .
\end{align}
\end{subequations}
This $Z_2$ interchanges the $\xh$ and $\yh$ axes of the daughter theory,
by interchanging the two $Z_N$ symmetries of the orbifold projection,
$\hat{\sigma}_d \hat{\gamma}_1\hat{\sigma}_d^{-1}=\hat{\gamma}_2$,
$\hat{\sigma}_d \hat{\gamma}_2\hat{\sigma}_d^{-1}=\hat{\gamma}_1$.  It
follows immediately that $\hat{\sigma}_d$ commutes with the orbifold
projection operator $\hat{P}$ as defined in \Eq{proj}.  The action of
$\hat{\sigma}_d$ on the lattice variables is
\begin{subequations}
  \eqn{zii}  \begin{align}
    x_{\mathbf{n}} & \to y_{\mathbf{\tilde{n}}} &\alpha_{\mathbf{n}}
    &\to \  \     \beta_{\mathbf{\tilde{n}}}\\
    \mybar x_{\mathbf{n}} &\to \mybar y_{\mathbf{\tilde{n}}}
    &\beta_{\mathbf{n}} &\to \ \ 
    \alpha_{\mathbf{\tilde{n}}}\\
    y_{\mathbf{n}} &\to x_{\mathbf{\tilde{n}}} &\lambda_{\mathbf{n}}
    &\to \ \ 
    \lambda_{\mathbf{\tilde{n}}}\\
    \mybar y_{\mathbf{n}} &\to \mybar x_{\mathbf{\tilde{n}}}
    &\xi_{\mathbf{n}} &\to -\xi_{\mathbf{\tilde{n}}}
\end{align}
\end{subequations}
where $\mathbf{n}=\{n_x,n_y\}\text{ and }\mathbf{\tilde n}=\{n_y,n_x\}$.

\subsection{The $\CQ=1$ supersymmetry and a superfield formulation}
\label{sec:4e}

Our method for constructing the lattice action for $(2,2)$ SYM in two
dimensions preserves an exact supersymmetry. Since the supersymmetry
transformations of the mother theory in \Eq{trans} are gauge
covariant, it follows that the four supercharges $Q_i$ of the mother
theory are gauge invariant.  Therefore the $\hat{\gamma}_a$ generators of
the $\Gamma$ symmetry used in the orbifold projection act on
the supercharges simply as $G_R$ transformations:
\begin{equation}
  \hat{\gamma}_a\, Q_i = \left(e^{2\pi i r_a /N}\right)_{ij} Q_j\ .
\end{equation}
Any linear combination of these four supercharges which have
$\mathbf{r}=\{0,0\}$ are invariant under $Z_N\times Z_N$ and survive the orbifold projection.

Consider the first of the transformations in \Eq{trans}
\begin{equation}
  \delta v_m = -i\mybar \psi\, \mybar\sigma_m\kappa + i \mybar\kappa
  \,\mybar \sigma_m\psi
\end{equation}
This is $G_R$ covariant if we consider the Grassmann spinor parameters
$\kappa$ and $\mybar \kappa$ as transforming under $G_R$ in the same
manner as $\psi$ and $\mybar \psi$ respectively. It follows that the
supercharges of the mother theory transform under the $G_R$
representation that is conjugate to that of the gauginos $\psi$ and
$\mybar \psi$, namely $(2,1)_{ +1/2} \oplus (1,2)_{-1/2}$.  Since one
of the four components of $\psi$ and $\mybar \psi$ has
$\mathbf{r}=\{0,0\}$---namely the component $\lambda$---one of the four
supercharges has charge $\mathbf{r}=\{0,0\}$ and survives the
orbifold projection.  This is a general result: for supersymmetric
lattices produced by orbifold projection from an SYM matrix model, the
number of unbroken supercharges equals the number of gaugino site
variables on the lattice.

To analyze the action of the unbroken supercharge, we write the
Grassmann spinors $\kappa$ and $\mybar \kappa$ which parameterize the
$\CQ=4$ supersymmetry transformations of the mother theory in terms of
the four independent components
\begin{equation}
  \kappa = 
  \begin{pmatrix}
    \eta  \cr \eta'\\
  \end{pmatrix} 
  \ , \qquad
  \mybar\kappa = 
  \begin{pmatrix}
    \mybar\eta  & \mybar\eta'
  \end{pmatrix}\ .
  \eqn{kap}
\end{equation}
With these definitions the upper component of $\kappa$, labeled as
$\eta$ in \Eq{kap}, transforms the same way as the spinor $\lambda$
under $G_R$ and has $\mathbf{r}=\{0,0\}$.  Therefore the supercharge
that generates supersymmetric translations in $\eta$ survives the
orbifold projection and gives rise to an exact supersymmetry of the
lattice action.

In order to study the renormalization properties of this lattice
theory, we rewrite the action in superfield notation to make the exact
$\CQ=1$ supersymmetry more explicit.  Setting $\eta'=\mybar
\eta=\mybar \eta'=0$ in \Eq{kap} and substituting into \Eq{trans}, we
find the supersymmetry transformations left unbroken by the orbifold
projection:
\begin{subequations}
  \eqn{pdfields}
\begin{align}
    \delta  x &=-\sqrt{2}\,i\,\alpha\,\eta    &
    \delta\mybar  x &= 0\\ 
    \delta  y&=-\sqrt{2}\, i\,\beta \,\eta  &
    \delta\mybar y&=0\\ 
    \delta\lambda &= -i\left([\mybar x, x]+[\mybar y, y]\right)\eta  
    &
    \delta\alpha&=0 \\
    \delta\xi &= 2i[\mybar x,\mybar y]\,\eta    &
    \delta\beta &= 0\ .
\end{align}
\end{subequations}
To construct a superfield notation that realizes the supersymmetry
algebra off-shell, we define the supercharge $Q$ in terms of a
Grassmann coordinate $\theta$:
\begin{equation}
  \delta = i\eta  Q\ ,\qquad Q=\frac{\partial\ }{\partial\theta}\ .
\end{equation}
The supersymmetry variations as written in \Eq{pdfields} are not
consistent with the definition of $\delta=i\eta \partial_\theta$ as
they stand, as $\delta^2\lambda\ne 0$.  We remedy this by introducing an
auxiliary field $d$ such that
\begin{align}
 \delta\lambda &= -i\left([\mybar x, x]+[\mybar y, y] + i
   d\right)\eta  \\
 \delta d &= 
 \sqrt{2}\left([\mybar x, \alpha]+[\mybar y, \beta]\right) \eta
\end{align}
which yields $\delta^2\lambda=0$. Making use of  the equations of
motion we will find $d=0$ and
$\delta d=0$.

With $\delta = i\eta\partial_\theta$, these transformation laws can be
realized in terms of the following superfields:
\begin{subequations}
  \eqn{sfields}
  \begin{align}
    \bfx &=  x +\sqrt{2}\,\theta \,\alpha\\
    \bfy &=  y+\sqrt{2}\,\theta \,\beta\\
    \boldsymbol{\Lambda} &= \lambda -\left([\mybar x, x]+[\mybar y, y] +i
      d\right)\,\theta\\
    \boldsymbol{\Xi} &= \xi + 2[\mybar x,\mybar y]\,\theta
  \end{align}
\end{subequations}
The action of the mother theory can be written in terms of these
superfields, as well as $\mybar x$ and $\mybar y$, which are
supersymmetric singlets (that is, matrices with no $\theta$
dependence).

After orbifolding, we label each surviving $k\times k$ block of the
original superfields by their lattice coordinate. The lattice
variables are the $\theta$-independent matrices 
$\mybar x_{\mathbf n}$ and $\mybar y_{\mathbf n}$, as well as the superfields
\begin{subequations}
  \eqn{sfields2}
  \begin{align}
    \bfx_{\mathbf n} &=  x_{\mathbf n} +\sqrt{2}\,\theta \,\alpha_{\mathbf n}\\
    \bfy_{\mathbf n} &=  y_{\mathbf n}+\sqrt{2}\,\theta \,\beta_{\mathbf n}\\
    \boldsymbol{\Lambda}_{\mathbf n} &= \lambda_{\mathbf n} -
    \left(({\mybar x}_{\mathbf{n}-\xh} x_{\mathbf{n}-\xh} -
      x_{\mathbf n}{\mybar x}_{\mathbf n}) +({\mybar
        y}_{\mathbf{n}-\yh} y_{\mathbf{n}-\yh} -y_{\mathbf
        n} {\mybar y}_{\mathbf n})  +i d_{\mathbf n}\right)\,\theta\\
    \boldsymbol{\Xi}_{\mathbf n} &= \xi_{\mathbf n} + 2\left({\mybar
        x}_{\mathbf{n}+\xh} {\mybar y}_{\mathbf n}- {\mybar
        y}_{\mathbf{n}+\yh} {\mybar x}_{\mathbf n}\right)\,\theta
  \end{align}
\end{subequations}
These superfields are all $k\times k$ matrices.  Note that the exact
lattice supersymmetry allows for variables which reside at different
places on the lattice to exist in the same super multiplet. For
example, $\boldsymbol{\Xi}_{\mathbf n}$ depends on $\xi_{\mathbf n}$,
which sits on the diagonal link, as well as the $\mybar x$ and $\mybar
y$ link variables residing at the edges of the plaquette that the
$\xi_{\mathbf n}$ link bisects.  In fact, the $\theta$ term of
$\boldsymbol{\Xi}$ resembles the square root of the usual plaquette
term found in lattice QCD.  It is this slightly nonlocal structure on
the lattice that allows successive supersymmetry transformations to
generate translations in the continuum limit.

In terms of these fields, the lattice action \Eq{q4lat} may be written
in a manifestly $\CQ=1$ supersymmetric form:
\begin{multline}
  S = \frac{1}{g^2} \int\! d\theta\,\sum_{\mathbf n} \Tr\biggl[-\half
  \boldsymbol{\Lambda}_{\mathbf n}\,\partial_\theta
  \boldsymbol{\Lambda}_{\mathbf n}  \\
  -\boldsymbol{\Lambda}_{\mathbf n}(\mybar
  x_{\mathbf{n}-\xh}{\bfx}_{\mathbf{n}-\xh} - {\bfx}_{\mathbf
  n}\mybar x_{\mathbf n}+\mybar y_{\mathbf{n}-\yh}
  {\bfy}_{\mathbf{n}-\yh}-{\bfy}_{\mathbf   n}\mybar y_{\mathbf n}
  ) \\  - \,\boldsymbol{\Xi}_{\mathbf n }\,({\bfx}_{\mathbf
  n}{\bfy}_{\mathbf{n}+\xh}  -  {\bfy}_{\mathbf n 
  }{\bfx}_{\mathbf{n}+\yh})\biggr] 
  \eqn{sfact}
\end{multline}
The lattice action \Eq{q4lat} may be recovered by integrating over the
auxiliary field $d_{\mathbf n}$. Note that $d_{\mathbf n}$ is a decoupled Gaussian
variable---its only function is to allow closure of the supersymmetry
algebra. 

This action has a $U(1)$ $R$-symmetry under which $\theta$,
$\lambda$ and $\xi$ have charge $+\half$, $\alpha$ and $\beta$ have
charge $-\half$, while $x$, $\mybar x$, $y$, and $\mybar y$ have
charge zero.  This is just the $U(1)_Y$ symmetry discussed in
\S~\ref{sec:4c}. In the continuum this becomes the vector $U(1)$
symmetry of the target theory \Eq{targ2}. Under the $Z_2$ crystal
symmetry \Eq{zii}, the superfields transform as
\begin{equation}
  \begin{aligned}
    \boldsymbol{\Lambda}_{\mathbf{n}} &\to \ \
      \boldsymbol{\Lambda}_{\mathbf{\tilde{n}}}\\
    \bfx_{\mathbf{n}} & \to  \ \ \bfy_{\mathbf{\tilde{n}}} &\qquad \mybar
    x_{\mathbf{n}} &\to\ \  \,\mybar y_{\mathbf{\tilde{n}}}\\
    \bfy_{\mathbf{n}} & \to  \ \  \bfx_{\mathbf{\tilde{n}}} &\qquad  \mybar
    y_{\mathbf{n}} &\to \ \ \, \mybar x_{\mathbf{\tilde{n}}}\\
    \boldsymbol{\Xi}_{\mathbf{n}} &\to
      -\boldsymbol{\Xi}_{\mathbf{\tilde{n}}} 
  \eqn{ziisf}
\end{aligned}
\end{equation}

\section{The continuum limit: tree level}
\label{sec:4}

The action \Eq{q4lat} has the ``topology'' of a lattice action, but
as yet no dimensionful  parameter identifiable as a lattice
spacing.  However, the action has a moduli space: flat
directions in the bosonic variables.  For example, the potential
vanishes for translationally invariant configurations of the form
$x=\mybar x$, $y=\mybar y$ and $[x,y]=0$. We consider the particular
configuration which preserves the $Z_2$ lattice symmetry and a $U(k)$
global symmetry, namely
\begin{equation}
  x_{\mathbf n} = \mybar x_{\mathbf n}=y_{\mathbf n} = \mybar
  y_{\mathbf n}= \frac{1}{a\sqrt{2}}\,{\mathbf 1}_k\ ,
\eqn{class}
\end{equation}
where $a$ is a parameter with dimensions of length, and ${\mathbf
  1}_k$ is the $k\times k$ unit matrix.  We expand the action about
this classical configuration \Eq{class}, identifying the lattice
spacing with $a$.  Then the continuum limit is taken as
\begin{equation}
  a\to 0\ ,\qquad N\to\infty\ ,\qquad g^2\to\infty\ ,\qquad g^2 a^2 =
  g_2^2\ ,\qquad a N = L\ ,
  \eqn{clim}
\end{equation}
where the two-dimensional gauge coupling $g_2$ and the box size $L$
are kept fixed. The infinite volume limit, $L\to \infty$ may be taken
separately. Because the target theory is superrenormalizable, the
dimensionless constant $g_2^2a^2$ vanishes in the continuum limit,
justifying a perturbative discussion of the continuum limit of this
model. All nonperturbative phenomena will be associated with
infrared physics.

To analyze the lattice \Eq{q4lat} we write $x$ and $y$ in terms of
their hermitean and anti-hermitean parts:
\begin{equation}
  x = \mybar x^\dagger = \frac{1}{a\sqrt{2}}+\frac{s_1 + i
  v_1}{\sqrt{2}}\ ,\qquad  y = \mybar y^\dagger =
  \frac{1}{a\sqrt{2}}+\frac{s_2 + i v_2}{\sqrt{2}}\ , 
\end{equation}
where both $s_i$ and $v_i$ are hermitean $k\times k$ matrices. Expanding about the point
in moduli space \Eq{class} and taking the limit \Eq{clim}, we show
below that the lattice
action \Eq{q4lat} becomes the action of the target theory \Eq{targ2}
where the continuum variables are defined in terms of the lattice
variables as
\begin{gather}
  s = \frac{s_1+is_2}{\sqrt{2}}\ ,\qquad
  v_m=\{v_1,v_2\}\ ,\qquad
  \psi = \begin{pmatrix} \lambda \cr \xi \end{pmatrix}\ ,\qquad
  \mybar\psi = i\begin{pmatrix}\alpha & \beta \end{pmatrix}\ ,
  \eqn{map}
\intertext{in the Dirac $\gamma$-matrix basis}
  \gamma_1=\sigma_3\ ,\quad \gamma_2=\sigma_1\ ,\quad
  \gamma_3 = \sigma_2\ .
  \eqn{dirac}
\end{gather}
The matrix $\gamma_3$ is the chirality matrix in two dimensions, with
$\psi_{R,L}$ in \Eq{targ2} defined as $\frac{1\pm \gamma_3}{2}\,
\psi$.
  
While simple to derive, the above result is startling.  For one thing,
the lattice has no discrete rotational symmetry---the $Z_2$ crystal
symmetry behaves more like parity---nevertheless an $SO(2)$ Euclidean
rotation symmetry is recovered in the continuum.  Furthermore, the
$s_{1,2}$ variables live on the links, leading to the natural
assumption that they transform nontrivially under spacetime rotations
in the continuum limit.  Instead $s= (s_1+is_2)/\sqrt{2}$ is a Lorentz
singlet in the continuum; the $SO(2)$ symmetry transforming $s_1$ and
$s_{2}$ into each other ends up being part of the internal chiral
$U(1)_R$ symmetry of the target theory, and not a spacetime symmetry.
In contrast, the anti-hermitean parts of the link variables $v_{1,2}$
indeed transform nontrivially under Lorentz rotations.  Finally, the
four one-component fermions, each located at different points on the
lattice, arrange themselves into a single Dirac spinor in the
continuum limit, with no extraneous doublers.

To determine the continuum limit of our lattice action, we begin by
considering the free kinetic terms at the classical level. After
eliminating the auxiliary field $d$ from the lattice action \Eq{sfact}
and expanding about the point in moduli space given in \Eq{class}, the
terms quadratic in the bosonic variables $s$ and $v$ are
\begin{align}
  \begin{split}
   \frac{1}{2 g^2 a^2}\, \sum_{\mathbf n}
    \Tr&\left[ \left(
        s_{1,\mathbf{n}-\xh} - s_{1,\mathbf n} +
    s_{2,\mathbf{n}-\yh} -  s_{2,\mathbf n} \right)^2
    \right. \\\notag
    &+\left.\Bigl\vert \left(s_{1,\mathbf{n} + \yh} - s_{1,\mathbf n} +
        s_{2,\mathbf n} - s_{2,\mathbf{n}+\xh}\right)
    -i\left(v_{1,\mathbf{n} + \yh} - v_{1,\mathbf n} -
    v_{2,\mathbf{n}+\xh}+  
        v_{2,\mathbf n}\right)\Bigr\vert^2 \right]
    \end{split}\\
    =\frac{1}{2 g^2}\, \sum_{\mathbf n}\Tr&\left[
     \sum_{\boldsymbol{\hat{\mu}}} \sum_{i=1,2}\left(\frac{s_{i,\mathbf n}-s_{i,\mathbf
            n-\boldsymbol{\hat{\mu}}}}{a}\right)^2 +
      \left(\frac{v_{1,\mathbf{n} + \yh} - v_{1,\mathbf n}}{a} -
        \frac{v_{2,\mathbf{n}+\xh} - v_{2,\mathbf
            n}}{a}\right)^2\right]\ , \eqn{bosii}
\end{align}
where $\boldsymbol{\hat{\mu}}$ is summed over $\xh$,
$\yh$.  We can diagonalize these kinetic terms by
introducing the Fourier transform of a field $\phi$
\begin{equation}
  \phi_{\mathbf n} \equiv \frac{1}{N} \sum_{\mathbf p} e^{i a {\mathbf p}\cdot
  {\mathbf n}}   \phi_{\mathbf p}
\end{equation}
where the momenta $p_i$ are integers times $\pi/L$, lying in the
Brillouin zone $p_i\in [\left.-\frac{\pi}{a},\frac{\pi}{a}\right)$.
The bosonic ``hopping'' terms \Eq{bosii} become
\begin{equation}
  \frac{1}{2g^2} \sum_{\mathbf p}\,\Tr\left[\left({\cal P}_x^2 +{\cal
        P}_y^2\right)  \left(s_{1,{\mathbf p}}s_{1,-{\mathbf p} }+s_{2,{\mathbf
          p}}s_{2,-{\mathbf p}}\right)  +
    \begin{pmatrix}
      v_{1,{\mathbf p}} & v_{2,{\mathbf p}}
    \end{pmatrix}
    G({\mathbf p})
    \begin{pmatrix}
      v_{1,-{\mathbf p}} \\ v_{2,-{\mathbf p}}
    \end{pmatrix}\right]
\end{equation} 
where we have defined
\begin{equation}
  {\cal P}_i \equiv
  \frac{2}{a}\sin\frac{ap_i}{2}\ ,\qquad
  {\mathbf G}({\mathbf p})\equiv 
  \begin{pmatrix}
    {\cal P}_y^2 &
    -e^{-ia(p_x-p_y)/2}{\cal P}_x{\cal P}_y\cr
    - e^{ia(p_x-p_y)/2}{\cal P}_x{\cal P}_y&
    {\cal P}_x^2\ .
  \end{pmatrix}
\eqn{vmodes}\end{equation}
The eigenvalues of ${\mathbf G}$ are $(\CP_x^2 + \CP_y^2)$ and zero.  In
the continuum limit ${\cal P}_i\to p_i$ and ${\mathbf G}({\mathbf p})_{ij}\to
({\mathbf p}^2\delta_{ij}-p_i p_j)$. We see that the continuum limit of
the boson lattice action correctly reproduces the kinetic
terms of the target theory
\begin{equation}
  \frac{1}{2g_2^2}\int \!d^2\!z\, \Tr \left[{\boldsymbol
      \nabla} s^\dagger\cdot {\boldsymbol \nabla} s
    + (\partial_x v_2 - \partial_y v_1)^2\right]\ ,
\end{equation}
where we used the definitions \Eq{map} and the substitution
$a^2\sum_{\mathbf n} \to \int \!d^2\!z$ as $a\to 0$.  Note that the above
Lagrangean has both a rotational symmetry under which $s$ is a scalar
and $\mathbf{v}$ is a vector, as well as a $U(1)$ symmetry under which
$s$ is charged and $v$ is neutral---neither symmetry being apparent in the lattice
action.

Following the same procedure for the fermionic part of the lattice
action \Eq{q4lat}, we find the noninteracting part of the fermion
hopping terms
\begin{gather} 
  \frac{1}{g^2} \sum_{\mathbf n} \Tr
  \begin{pmatrix}
    \alpha_{\mathbf p} & \beta_{\mathbf p}
  \end{pmatrix}
  i {\mathbf K}({\mathbf p})
  \begin{pmatrix}
    \lambda_{-\mathbf p} \cr \xi_{-\mathbf p}
  \end{pmatrix}
\intertext{with the definition}
  {\mathbf K}({\mathbf p}) =\begin{pmatrix} 
    e^{-ia p_x/2}\CP_x & e^{iap_y/2}\CP_y\cr
    e^{-iap_y/2} \CP_y & -e^{iap_x/2}\CP_x
  \end{pmatrix}
\end{gather} 
The matrix ${\mathbf K}^\dagger \mathbf{K}$ is just the identity
matrix times $(\CP_x^2 + \CP_y^2)$, which equals the scalar hopping
matrix, as expected on a supersymmetric lattice. Note that there is no
fermion doubling at the edges of the Brillouin zone, $a p_i =\pm \pi$.
and the continuum limit of the free part of the fermion action
is just the two dimensional Dirac action
\begin{equation}
  \frac{1}{g_2^2}\int\! d^2\!z\, \Tr \mybar \psi
  \,i\dslash \psi\ .
\end{equation}
with $\psi$ as defined in \Eq{map}, and $\gamma$ matrices defined in
\Eq{dirac}. This has the same form as in the target theory.

The above analysis has shown that only smooth configurations (up to
gauge transformations) have small action, and we are justified in
studying the continuum limit of the interacting theory by expanding
about smooth field configurations, limiting ourselves to smooth gauge
transformations.  To retain manifest $\CQ=1$ supersymmetry, we define
continuum versions of the lattice superfields in \Eq{sfields}; these
are functions of $\mathbf{z}$, the spacetime coordinate in two
Euclidean dimensions:
\begin{alignat}{2}
  -\frac{i}{\sqrt{2}}(\mathbf X_{\mathbf n} -\mybar x_{\mathbf n}) \ 
  &\to\quad & \mathbf V_1(\mathbf z) &= v_1(\mathbf z) -i\theta
  \,\alpha(\mathbf  z)\notag\\
  -\frac{i}{\sqrt{2}}(\mathbf Y_{\mathbf n} -\mybar y_{\mathbf n}) \ &\to &
  \mathbf V_2(\mathbf z) &= v_2(\mathbf z) -i\theta
  \,\beta(\mathbf z)\notag\\
  \frac{1}{\sqrt{2}}(\bfx_{\mathbf n} +\mybar x_{\mathbf n}) -\frac{1}{a}\ &\to &
  \bfs_1(\mathbf z) &= s_1({\mathbf z}) +\theta\alpha(\mathbf
  z)\eqn{smooth}\\
  \frac{1}{\sqrt{2}}(\bfy_{\mathbf n} +\mybar y_{\mathbf n}) -\frac{1}{a}\ &\to &
  \bfs_2(\mathbf z) &= s_2(\mathbf z) +\theta\beta(\mathbf z)\notag\\
  \phantom{\frac{1}{\sqrt{2}}}\boldsymbol{\Lambda}_{\mathbf n} \ &\to &
  \boldsymbol{\Lambda}(\mathbf z) \, &= \lambda(\mathbf z) + \theta \left(D_1
    s_1(\mathbf z) + D_2
    s_2(\mathbf z) - id(\mathbf z)  + O(a) \right)\notag\\
  \phantom{\frac{1}{\sqrt{2}}}\mathbf\Xi_{\mathbf n} \ &\to &
  \boldsymbol{\Xi}(\mathbf z)\, &= \xi(\mathbf z) + \theta\left(D_2 s_1(\mathbf z)
    - D_1 s_2(\mathbf z)+[s_1,s_2]+ i v_{12}(\mathbf z)+ O(a) \right)\ 
  ,
\notag
\end{alignat}
where $D_{1,2}$ are the gauge covariant derivatives, $D_m=\partial_m +
i [v_m,\;\cdot\; ]$, and $v_{mn}$ is the gauge field strength. Note
that the Grassmann superfields $\boldsymbol{\Lambda}$ and
$\boldsymbol{\Xi}$ have $O(a)$ corrections.

Consider how the continuum fields transform under the lattice
symmetries. Under smooth gauge transformations, all of these
superfields transform homogeneously as $\Phi\to U \Phi U^\dagger$,
with the exception of the gauge field multiplet which transforms in
the usual way as $\bfv_m\to U\bfv_m U^\dagger + i (\partial_m U)
U^\dagger$.  By specifying ``smooth'' gauge transformations, we mean
that the transformation laws have $O(a)$ corrections, which may be
computed from \Eq{smooth}.  The unbroken global $U(1)$ symmetry---the
$U(1)_Y$ symmetry of the mother theory---acts as a $U(1)_R$ symmetry
on these fields, with $\boldsymbol{\Lambda}$, $\boldsymbol{\Xi}$ and
the Grassmann coordinate $\theta$ carrying charge $-\half$, while
$\bfv_m$ and $\bfs_i$ have charge zero. Finally, the $Z_2$ crystal
symmetry interchanges the fields $\bfv_1\leftrightarrow \bfv_2$ and
$\bfs_1\leftrightarrow\bfs_2$, while $\boldsymbol{\Xi}\to
-\boldsymbol{\Xi}$ and $\boldsymbol{\Lambda}\to \boldsymbol{\Lambda}$;
the coordinate also changes under the $Z_2$, with $\mathbf{z}\to
\mathbf{\tilde z}$, where $\mathbf{z}=\{x_1,x_2\}$ and
$\mathbf{\tilde{z}} = \{x_2,x_1\}$.

In terms of these continuum $\CQ=1$ superfields, the lattice action
\Eq{sfact} has the continuum expansion
\begin{multline}
  S = \frac{1}{g_2^2}\int\! d\theta\!\int \!d^2\!z\,\Tr \biggl[ -\half
  \boldsymbol{\Lambda} \partial_\theta \boldsymbol{\Lambda} +
    \boldsymbol{\Lambda}\left(\CD_1{\bfs_1} + \CD_2\bfs_2 \right) \\
  -\half \boldsymbol{\Xi} \bigl( \CD_1\bfs_2 - \CD_2\bfs_1 +
  [\bfs_1,\,\bfs_2] + i \bfv_{12}\bigr)\biggr]+ O(a)\ , \eqn{contact}
\end{multline}
where we have defined a supersymmetric covariant derivative and field
strength,
\begin{equation}
  \CD_m = \partial_m + i\bigl[ \bfv_m,\;\cdot\; \bigr]\ ,\qquad
  \bfv_{mn}= -i\left[ \CD_m,\,\CD_n\right]\ .
\end{equation}
It is not difficult to verify in component form that the above
action agrees  with the action of the target theory \Eq{targ2}.

The analysis up to this point has been purely classical.  We now
discuss renormalization and show that radiative effects do not spoil
the nice features we found at tree level.

\section{The continuum limit: renormalization}
\label{sec:5}

We turn now to the quantum theory. At tree level we have the desired
continuum limit, but radiative effects could in principle induce
unwanted relevant or marginal operators which do not respect the
symmetries of the target theory.  Following the analysis in
ref.~\cite{Kaplan:2002wv} we consider the lattice action expanded
about the classical continuum limit in powers of the lattice spacing
$a$. We then consider the possible operators that might be required as
counterterms in a loop expansion, according to standard power counting
arguments.  Since the two-dimensional gauge coupling constant $g_2$
has mass dimension 1, the loop expansion is governed by powers of
$g_2^2a^2$, which vanishes in the continuum limit \Eq{clim},
justifying a perturbative analysis. The allowed counterterms are
restricted by the exact symmetries of the lattice theory.  We show
that the only possible counterterm allowed by these symmetries is a
shift in the vacuum energy, which does not affect the physics of the
target theory in the continuum limit.

Consider the radiative addition to the action of  an operator $\CO$ 
\begin{equation}
  \delta S=\frac{1}{g_2^2} \int\! d\theta\! \int\! d^2\!z\ C_{\CO}\,\CO\ .
\end{equation}
$\CO$ must be Grassmann with $U(1)_R$ charge equal to $-\half$, and is
constructed from the superfields, their $\theta$-derivatives and
covariant derivatives.  For power counting purposes, $\int\! d^2\!z$
has dimension $-2$, $\theta$ has dimension $-\half$, and $\int
d\theta$ and $\partial_\theta$ scale with dimension $+\half$.  A
contribution to $C_{\CO}$ at $\ell$ loops is proportional to
$g_2^{2\ell}$.  Since the coupling $g_2$ has mass dimension 1, the
loop expansion is an expansion in the dimensionless parameter $(g_2^2
a^2)^\ell$.  Consequently the coefficient $C_{\CO}$ of the operator
$\cal O$ scaling with mass dimension $p$ has a loop expansion
\begin{equation}
  C_{\CO} = a^{p-7/2}\,\sum_\ell c_\ell\, (g_2^2 a^2)^\ell\ ,
\end{equation}
where the dimensionless expansion coefficients $c_\ell$ depend at most
logarithmically on the lattice spacing $a$.

The only possible local counterterms (operators whose coefficients do
not vanish in the $a\to 0$ limit) are those for which $(p+2\ell-7/2)
\le 0$.  We have shown that at tree level ($\ell=0$) the effective
action of the lattice theory agrees with that of the target theory,
and so we need only consider $\ell\ge 1$, and hence $0\le p\le 3/2$.

The superfield dimensions may be determined by the natural scaling of
the lowest component, so the bosonic superfields $\mathbf{S} \text{
  and } \mathbf{V}$ have dimension 1, while the Grassmann superfields
$\boldsymbol{\Lambda} \text{ and } \boldsymbol{\Xi}$  scale
with mass dimension $\frac{3}{2}$.  Since the operator $\CO$ must be
Grassmann, our power counting arguments imply it must have the form
$\Tr \partial_\theta \mathbf{B}$, where $\mathbf{B}$ is a bosonic
superfield, or $\Tr \boldsymbol{\Gamma}$, where $\boldsymbol{\Gamma}$
is a Grassmann superfield.  The requirement that $\CO$ carry $U(1)_R$
charge $-\half$ precludes the former possibility, while the $Z_2$
crystal symmetry forbids $\CO=\Tr \boldsymbol{\Xi}$.  The only
possible counterterm is therefore
\begin{equation}
  \label{eq:1}
  \frac{1}{g_2^2} \int\! d\theta\! \int\! d^2\!z\
  i\xi\,\Tr\boldsymbol{\Lambda} = \frac{1}{g_2^2} \int\! d^2\!z\  \xi
  \,\Tr d\ . 
\end{equation}
where we have dropped integrals of total derivatives.  This is a
Fayet-Illiopoulos term for the $U(1)$ part of the $U(k)$ gauge group.

How does this term affect the theory? Since $d$ enters the action as a
free, Gaussian field (recall that the $d$ equation of motion is
$d=0$), this term shifts the value of $d$ to a nonzero constant, which
in turn merely adds a constant to the vacuum energy.  After
integrating $d$ out of the theory, all that is left is a cosmological
constant which has no affect on the spectrum nor on the interactions
of the continuum limit of the theory.

We conclude that the target theory \Eq{targ2} is obtained from the
continuum limit of our lattice action without any fine tuning, modulo
a possible uninteresting contribution to the vacuum energy density.

\section{Moduli}
\label{sec:6}

The expansion of the lattice fields about the point
\begin{equation}
  x_{\mathbf n} = \mybar x_{\mathbf n}=y_{\mathbf n} = \mybar
  y_{\mathbf n}= \frac{1}{a\sqrt{2}}\,{\mathbf 1}_k\ ,
  \eqn{classb}
\end{equation}
in the classical moduli space of the lattice theory requires
justification. In particular, quantum fluctuations about this vacuum
must be small compared to the classical value.  Before considering the
lattice theory it is necessary to discuss the moduli of the continuum
target theory.  The action \Eq{targ2} possesses a noncompact classical
moduli space consisting of all constant scalar field configurations
$s$ satisfying $[s,s^\dagger]=0$.  The existence of these exact
noncompact bosonic zeromodes in the target theory poses an apparent
problem for the lattice theory.  The integrals over these modes are
formally divergent, and the expansion about \Eq{classb} is then poorly
defined.

To eliminate the noncompact modes associated with any diagonal matrix
value for $s$, up to gauge transformations, we add to the target
theory a small mass term $\mu^2 \Tr |s|^2$, which lifts the degeneracy
of the moduli and favors $s=0$.  This term breaks supersymmetry softly
(but not the gauge symmetry), and will be sent to zero as $1/L$ in the
infinite volume limit.  Physical questions on length scales smaller
than $1/\mu$ (and hence smaller than the lattice size $1/L$) are
unaffected by this mass term.

Eliminating the zeromodes from the target theory might seem
insufficient---after all, the spontaneous breaking of compact
symmetries cannot occur in two dimensions even if the zero mode of the
order parameter is eliminated, due to the infrared divergences of the
momentum $k\ne 0$ modes of the Goldstone bosons
\cite{Mermin:1966fe,Coleman:1973ci}. Such is not the case here.
Consider the correlator of the scalar field $s$ in the continuum:
\begin{equation}
  G(\mathbf{z_1, z_2}) = \bigl\vert \langle s^\dagger(\mathbf{z_1})
  s(\mathbf{z_2})\rangle\bigr\vert . 
\end{equation}
Assuming that the mass term we add for $s$ satisies $\mu\lesssim 1/L$,
the contribution to $G(\mathbf{z_1}, \mathbf{z_2})$ from nonzero
momentum modes is given by
\begin{equation}
  G(\mathbf{z_1, z_2})\Bigr\vert_{k\ne 0} \sim g_2^2 \ln
  \frac{|\mathbf{z_1}-\mathbf{z_2}|}{L} \lesssim  g_2^2 
  \ln \frac{a}{L}\ ,
\eqn{knz}
\end{equation}
where we made explicit that the target theory is regulated in the
ultraviolet by the lattice spacing $a$.  We see that the root mean
square fluctuations of $s$ diverges in the infinite volume limit. This
is a physical effect that cannot be eliminated from the theory.
However, these infrared divergent fluctuations need not invalidate the
expansion of the link variables about the classical values
\Eq{classb}, since the point we are expanding about corresponds to the
ultraviolet cutoff, and not a physical scale, as would be the case for
spontaneous symmetry breaking in two dimensions.  So long as we take
the continuum limit of the lattice theory such that
\begin{equation}
  a^2 g_2^2\ln L/a \to 0\ ,
\end{equation}
the fluctuations of the modulus in the continuum theory do not destroy
the lattice interpretation of our orbifold construction.

Now that we understand the moduli of the continuum theory, all that
remains is to specify the modifications of the lattice action which
implement these changes in the continuum.  
We introduce a small supersymmetry violating term in the
lattice action to stabilize the noncompact bosonic zeromodes
\begin{equation}
  \delta S = -\frac{1}{g^2} \int d\theta \sum_{\bf n} \Tr \left[
    a^2\,  {\mathbf F} \left[ \left( {\bfx_{\bf n}}\mybar x_{\bf n}
        -\frac{1}{2a^2}\right)^2 + \left( {\bfy_{\bf n}}\mybar y_{\bf n}
        -\frac{1}{2a^2}\right)^2\right]\right]\ ,
\end{equation}
where $\mathbf{F}$ is a supersymmetry breaking spurion of the form
\begin{equation}
 \mathbf{F} = \theta \mu^2
\end{equation}
and $\mu^2$ is positive and proportional to the unit matrix.  The
addition of this term does not alter the UV properties of the theory.
Expanding this new term about the continuum fields of \Eq{smooth} leads
to an addition to the action
\begin{equation}
  \delta S =  -\frac{1}{g_2^2}\int\! d\theta\!\int \!d^2\!z\,\Tr
  {\mathbf F} \left({\mathbf S}_1^2 +{\mathbf S}_2^2\right)+ O(a) =
  -\frac{1}{g_2^2}\int \!d^2\!z\, \mu^2\Tr |s|^2+ O(a)\ . 
  \eqn{ds} 
\end{equation}
Classically this new term lifts the degeneracy of the vacua, giving
the would-be moduli a positive mass squared $\mu^2$ about the
classical value $x=y=1/a\sqrt{2}$, or equivalently, $s=0$.  As
discussed above, we can take $\mu\sim 1/L=1/Na$, and this mass term
alters the vacuum of the continuum theory in an irrelevant way for
correlators insensitive to mass scales of order $1/L$.

\section{Discussion}
\label{sec:8}
We have shown how to construct a supersymmetric Euclidean spacetime
lattice.  The present example has as its target theory $(2,2)$
supersymmetry in two dimensions, but as was the case in
Ref.~\cite{Kaplan:2002wv}, the method is generalizable to theories
with more supersymmetry and in higher dimensions.  A detailed
discussion of lattice theories possessing continuum limits with eight
or sixteen conserved supercharges is in preparation. One could also
try to generalize the four supercharge theory discussed here to
include matter fields in the fundamental representation of the gauge
group.

The present example does not require fine tuning, due to the
underlying symmetries of the lattice.  This is remarkable not only
because the target theory is supersymmetric, but also because it
possesses a chiral symmetry which is conserved up to anomalies.
Nevertheless, practical obstacles will have to be overcome if the
theory is to be simulated.  In general, a dynamical simulation with
light fermions (of mass $\lesssim 1/L$, where $L$ is the lattice size)
is required in order to obtain a SYM theory in the continuum limit.
In addition, even though the target theory has a positive fermion
determinant in the present example, the lattice theory presented here
may not have.  An interesting path to pursue which we have not
explored would be to construct the Nicolai map
\cite{Nicolai:1980nr,Nicolai:1980jc} for our lattice theory---this
could in principle circumvent the usual problems with simulating
dynamical fermions\footnote{We thank M.  L\"uscher for making this
  suggestion. Previous references discussing the Nicolai map in the
  context of lattice supersymmetry include
  \cite{Sakai:1983dg,Kikukawa:2002as}.}.

\acknowledgments We would like to thank A. Karch, M. L\"uscher, S.
Sethi, M. Strassler, and C. Vafa for conversations and correspondence.
D.B.K. and M.U. were supported in part by DOE grant DE-FGO3-00ER41132,
E.K by DOE grant DE-FG03-96ER40956, and A.G.C. by DOE grant
DE-FG03-91ER-40676. We would like to thank the Aspen Center for
Physics, where some of this work was done.

\bibliography{latticeSUSY2}
\bibliographystyle{JHEP} 

\end{document}